\documentclass[12pt]{iopart}

\usepackage{iopams} 
\usepackage{subfigure,graphicx}
\usepackage{pgf,tikz,tikz-cd}
\usetikzlibrary{calc, shapes, intersections,shapes.arrows,arrows.meta,decorations.markings,decorations.pathreplacing,angles,quotes}

\usepackage{cite}
\usepackage[hyphens]{url}
\usepackage   [ colorlinks=false,    
    pdfborder={0 0 0},   
    hidelinks            
    ]{hyperref}
\begin{document}

\title[S Verma et al]{Design of a three-dimensional parallel-to-point imaging system based on inverse methods}

\author{Sanjana Verma$^{{1,\star}}$, Lisa Kusch$^{1}$,  Koondanibha Mitra$^{1}$, Martijn J.H. Anthonissen$^{1}$, Jan H.M. ten Thije Boonkkamp$^{1}$, and Wilbert L. IJzerman$^{2,1}$}

\address{$^1$ Department of Mathematics and Computer Science, Eindhoven University of Technology, PO Box 513, 5600 MB Eindhoven, The Netherlands\\
		 $^2$ Signify Research, High Tech Campus 7, 5656 AE Eindhoven, The Netherlands}
\ead{$^{\star}$ s.verma@tue.nl}
\vspace{10pt}

\begin{abstract}
 We present an inverse method for designing a three-dimensional imaging system comprising of freeform optical surfaces. We impose an imaging condition on the optical map and combine it with the law of conservation of energy to conclude that the ratio of energy distributions at the source and target of an imaging system must be constant. A mathematical model for the design of a parallel-to-point system consisting of two freeform reflectors is presented. A Schwarzschild telescope, a classical design known for maximum correction of third-order aberrations, is utilized to specify the design parameters in the mathematical model, enabling us to compute an inverse freeform imaging system. The performance of both designs is compared by ray tracing various parallel beams of light and determining the corresponding spot sizes of the image. We demonstrate that our inverse freeform design is superior to the classical design.
\end{abstract}

\vspace{2pc}

\section{Introduction}\label{sec: intro}


 In an ideal image, all rays emitted from a source point meet at exactly one point on the image plane. Imaging systems are designed with the goal of minimizing optical aberrations, which are the deviations from an ideal image. Imaging optical design has traditionally relied on so-called \textit{forward} methods, in which the system is designed based on analytical methods, such as aberration theory \cite{korsch}, and its performance is predicted through simulations, like ray tracing. Subsequently, it is iteratively refined until the performance criteria are met. These methods are time-consuming and can become complicated for systems with a large number of optical surfaces \cite{braat}.

Due to advancements in fabrication techniques, the use of imaging systems with freeform optical surfaces, which are characterized by the absence of any symmetry, is becoming prominent. Recent methods for designing such systems involve optimization methods and nodal aberration theory \cite{braat,opticaldesignsoftware,evolutionary,NAT}. The existing methods depend on the experience of the optical designer and pose challenges like determining a good initial guess \cite{evolutionary,braat}. 

 In contrast to forward methods, \textit{inverse} methods aim to design systems from a specified performance goal. 
In our prior work \cite{sanjana}, we demonstrated that optical systems designed using inverse methods from \textit{nonimaging} optics \cite{lotte} can be optimized for aberration compensation, offering insight into the feasibility of adapting these methods to imaging systems. We aim to develop inverse methods for \textit{imaging} optics based on mathematical models from nonimaging optics. 
 
Inverse methods in nonimaging optics determine the optical surfaces by finding an optical map using the principles of geometrical optics and the law of conservation of energy \cite{lotte}. The energy distributions at the source and target are the input parameters, and the optical map is dependent on their ratio. In contrast, for an imaging system, the optical map must conform to first-order optics, i.e., paraxial optics \cite{hecht}. In other words, for the inverse design of an imaging system, we specify our desired optical map. To utilize inverse methods from nonimaging optics for designing imaging systems, we impose an imaging condition on the optical map and determine the required energy distributions for imaging design. In this paper, the object and image planes of an imaging system are referred to as the source and target planes, respectively, for consistency of nomenclature with nonimaging optics.

In \cite{sanjana2D}, we presented a proof-of-concept using the aforementioned approach to design a two-dimensional (2D) parallel-to-point imaging system with two freeform reflectors. In this paper, we extend it to design a three-dimensional (3D) double-reflector imaging system. An imaging system consisting of a parallel source and a point target is a telescopic system. Studies show that two-mirror telescopes with rotational symmetry correct multiple types of aberrations \cite{korsch,telescopeSasian}. The optical surfaces that we compute in nonimaging optics are freeform, implying that the imaging design based on inverse methods will also have freeform surfaces. However, we will restrict ourselves to rotationally symmetric systems in order to reduce the complexity of the mathematical model. This enables us to determine the optical map essential for ideal imaging in our optical system. The freeform design is verified using a highly accurate ray tracer based on B-spline quasi-interpolation \cite{QI,QIlecturenotes,QIlychee}. 
 
The structure of this paper is as follows. In Sec.~\ref{sec: imaging}, the condition for a phase-space optical map to produce zero aberrations is discussed. The mathematical model for the design of a 3D parallel-to-point double-reflector imaging system is presented in Sec.~~\ref{sec: model}. In Sec.~\ref{sec: inverse imaging}, we discuss the design procedure and elaborate that the input parameters depend on a classical design. The performance of the freeform design found by inverse methods is compared to the classical design in Sec.~\ref{sec: results}, followed by some conclusions and proposed extensions in Sec.~\ref{sec: conclusion}.  

\section{Imaging condition}\label{sec: imaging}
In geometrical optics, a light ray is determined by its
position vector $\boldsymbol{\vec{q}} = (q_1,q_2,z)^{\rm{T}}$ and its momentum vector
$\boldsymbol{\vec{p}} = (p_1,p_2,p_{z})^{\rm{T}}$, i.e., its direction cosines times the refractive
index $n$ of the medium. We will use the symbol $(\,\,\boldsymbol{\vec{}}\,\,)$ to denote 3D vectors. We will focus on reflective optical systems, so $n=1$. Let the $z$-axis be the optical axis, and the $z$-coordinate be the evolution parameter of a ray. The phase-space variables denoted by $(\boldsymbol{q},\boldsymbol{p})$, where $\boldsymbol{q}=(q_1,q_2)^{\rm{T}}\in\mathbb{R}^2$ and $\boldsymbol{p}=(p_1,p_2)^{\rm{T}}\in\mathbb{R}^2$, completely describe a ray at a plane $z= \rm{constant}$. Furthermore, the definition of momentum vector implies $|\boldsymbol{\vec{p}}|=1$. The third component $p_{{z}}$ of the momemtum vector $\boldsymbol{\vec{p}}$ is then given by
\begin{equation}\label{eq: 3D pz}
p_{{z}}=\sigma \sqrt{1-|\boldsymbol{p}|^2},
\end{equation}
where the variable $\sigma$ corresponds to forward $(\sigma=1)$ or backward $(\sigma=-1)$ propagation of the rays, when $p_{{z}}>0$ or $p_{{z}}<0$, respectively.

The optical Hamiltonian $H(\boldsymbol{q},\boldsymbol{p})$ governs the evolution of the variables $(\boldsymbol{q},\boldsymbol{p})$ as a ray propagates along the $z$-axis. A Hamiltonian system reads
\numparts
\begin{equation}\label{eq: 3D hamiltonian}
\boldsymbol{q}'=\frac{\partial H}{\partial \boldsymbol{p}},\quad  \boldsymbol{p}'=-\frac{\partial H}{\partial \boldsymbol{q}},\end{equation}
 \begin{equation}
 H(\boldsymbol{q},\boldsymbol{p})=-\sigma\sqrt{1-\left|\boldsymbol{p}\right|^2}=-p_{{z}},
 \label{eq: 3D hamiltonian defn}
\end{equation}
\endnumparts
where $\boldsymbol{q}'$ and $\boldsymbol{p}'$ are the rates of change of $\boldsymbol{q}$ and $\boldsymbol{p}$ with respect to the $z$-coordinate. 

Let the light rays emitted from a source plane have phase-space coordinates $(\boldsymbol{q}_{\rm{s}},\boldsymbol{p}_{\rm{s}})$. They pass through an imaging optical system and reach the target (image) plane with coordinates $(\boldsymbol{q}_{\rm{t}},\boldsymbol{p}_{\rm{t}})$. In other words, the source coordinates are transformed to the target coordinates. This transformation from the initial phase-space coordinates at the source to the final phase-space coordinates at the target is governed by the optical Hamiltonian in Eq.~(\ref{eq: 3D hamiltonian}). In the remainder of this paper, we use the subscripts $\rm{s}$ and $\rm{t}$ to denote source and target, respectively. We denote the optical map connecting these source and target coordinates by $\mathcal{M}:(\boldsymbol{q}_{\rm{s}}, \boldsymbol{p}_{\rm{s}}) \mapsto(\boldsymbol{q}_{\rm{t}},\boldsymbol{p}_{\rm{t}})$, with
\begin{equation}\label{eq: 3D phase-space optical map general}
 \left[ \begin{array}{cc}\boldsymbol{q}_{\rm{t}}\\ \boldsymbol{p}_{\rm{t}}\end{array}\right]=\mathcal{M}\left[\begin{array}{cc}\boldsymbol{q}_{\rm{s}}\\\boldsymbol{p}_{\rm{s}}\end{array}\right]=\left[\begin{array}{cc} M_1(\boldsymbol{q}_{\rm{s}},\boldsymbol{p}_{\rm{s}})\\M_2(\boldsymbol{q}_{\rm{s}},\boldsymbol{p}_{\rm{s}})\end{array}\right].
\end{equation}

The Jacobian matrix $M$ of the optical map $\mathcal{M}$ satisfies \cite[p.~164]{lagrangian}
\numparts
\begin{equation}\label{eq: sympletic}
MJM^{\rm{T}}=J,
\end{equation}
\begin{equation} 
J=\left[\begin{array}{cc}0&I_{2\times2}\\
-I_{2\times2}&0\end{array}\right].\label{eq: J matrix}
\end{equation}
\endnumparts
where $J$ is the antisymmetric matrix and $I_{2\times2}$ denotes the identity matrix with dimension $2\times2$. The relation in Eq.~(\ref{eq: sympletic}) means that $M$ is a symplectic matrix and $\det(M)=1$ \cite{lagrangian}. This implies that the optical map $\mathcal{M}$ is symplectic, meaning that the phase-space transformation of the light beam from the source to the target is volume-preserving \cite[p.~564]{symplecticarea}. 

Lie algebraic methods can be utilized to express the symplectic map $\mathcal{M}$ as a concatenation of Lie transformations. This holds under the condition that the origin in phase space is mapped to itself, i.e., $\mathcal{M}(\boldsymbol{0})=\boldsymbol{0}$. The aforementioned concatenation acts on the phase-space coordinates at the source and yields an expression consisting of linear and non-linear parts. The linear part corresponds to paraxial optics and is free of all aberrations \cite{aberrationfree}.  On the other hand, the non-linear parts comprise the aberrations produced by the optical system. A detailed explanation of the Lie algebraic methods is beyond the scope of this work.

Our aim is to design an optical system with minimal aberrations. The optical map $\mathcal{M}$ for an ideal imaging system should be linear. This implies that the optical map $\mathcal{M}$ is given by

\begin{equation}\label{eq: 3D linear matrix for imaging}
  \left[  \begin{array}{cc} \boldsymbol{q}_{\rm{t}} \\ \boldsymbol{p}_{\rm{t}} \end{array}\right]= \left[\begin{array}{cc} \mathcal{A}&\mathcal{B}\\ \mathcal{C}&\mathcal{D}\end{array}\right]\left[\begin{array}{cc} \boldsymbol{q}_{\rm{s}} \\ \boldsymbol{p}_{\rm{s}} \end{array}\right],
\end{equation}
where $\mathcal{A}$,  $\mathcal{B}$, $\mathcal{C}$, and $\mathcal{D}$, are constant matrices of dimensions $2\times2$.
Since Eq.~(\ref{eq: 3D linear matrix for imaging}) defines a linear optical map $\mathcal{M}$ and consists of constant block matrices, it follows that
\begin{equation}\label{eq: corresponding ABCD}
 \frac{\partial \boldsymbol{q}_{\rm{t}}}{\partial \boldsymbol{q}_{\rm{s}}}=\mathcal{A}, \quad \frac{\partial \boldsymbol{q}_{\rm{t}}}{\partial \boldsymbol{p}_{\rm{s}}}=\mathcal{B}, \quad
 \frac{\partial \boldsymbol{p}_{\rm{t}}}{\partial \boldsymbol{q}_{\rm{s}}}=\mathcal{C},\quad
 \frac{\partial \boldsymbol{p}_{\rm{t}}}{\partial \boldsymbol{p}_{\rm{s}}}=\mathcal{D}.
\end{equation}

\section{Mathematical model}
\label{sec: model}
\begin{figure}[htbp]
\centering
\begin{tikzpicture}[scale=0.8]
	\draw  [->] (-1.5,-1.5) node [left] { $z=-l$} -- (7,-1.5) node [below] {$x_2$};
				\draw [->](-1.5,4.5) node [left] {$z=0$} -- (7,4.5)node [below] {$y_2$};
				\draw  [->] (5,-2) -- (5,6) node [left] {$z$};	
				\draw [black,line width=2.2pt,name path=reflector1,text width=3.5cm] (-1,0.8) arc(140:110:5) node [left, blue,yshift=-0.7cm,xshift=-1.3cm,align=right] {$\mathcal{R}_1$: $z=-l+u(\boldsymbol{x})$};
				\draw [blue, ultra thick] (-1,-1.5) -- (1.2,-1.5) node [below] {{$\mathcal{S}$}};
				
				\draw [black,line width=2.2pt,name path=reflector2,text width=3cm] (2.8,-0.46) arc(270:330:4.1) node [below right, yshift=-0.7cm,xshift=-0.7cm, blue] {$\mathcal{R}_2$: $r=-w\big(\mathbf{\hat{\vphantom{X}t}}\big)\!\cdot{\mathbf{\hat{\vphantom{X}t}}}$};
				\coordinate (T) at (5,4.5);
				
				\def\arrowpos{0.6};
				\path [name path=rayin] (0.5,0) -- (0.5,5);
				\begin{scope}[decoration={markings,mark=at position \arrowpos with {\arrow{stealth}}}]
				\path [name intersections={of=rayin and reflector1, by=A}];
				\draw [orange, ultra thick, postaction={decorate}] (0.5,-1.5) -- (A)  node [midway, left] {${\mathbf{\hat{\vphantom{X}s}}}$};
				\path [name path=rayout] (3.5,-2) -- (T);
				\path [name intersections={of=rayout and reflector2, by=B}];

                \def\arrowpos{0.5};
				\draw [orange, ultra thick, postaction={decorate}] (A) -- (B) node [midway,below left,xshift=0.1cm] {${\mathbf{\hat{\vphantom{X}i}}}$};
				\def\arrowpos{0.55};
				\draw [orange, ultra thick, postaction={decorate}] (B) -- (T) node [midway, left] {${\mathbf{\hat{\vphantom{X}t}}}$};
				\end{scope}	
				\draw (A) node [above,yshift=1mm] {$P_1$};
				\draw (B) node [below] {$P_2$};
				\path [name path=lastpartray] (B) -- (T);
				\path [name path=unitcircle] (T) circle(1cm);
				\path [name intersections={of=lastpartray and unitcircle, by=C}];
				\filldraw [blue] (T) node [above right] {{$\mathcal{T}$}} circle(3pt);

                \definecolor{mydarkgray}{RGB}{90,90,90} 
                \begin{scope}[shift={(10,-1.5)}, rotate=0]
               \draw[->,thick, mydarkgray] (0,0,0) -- (1,0,0) node[right] {$x_2$}; 
             \draw[->,thick, mydarkgray] (0,0,0) -- (0,1,0) node[above] {$z$}; 
              \draw[->,thick, mydarkgray] (0,0,0) -- (0,0,1.5) node[below left] {$x_1$}; 
              \end{scope}
               \begin{scope}[shift={(10,4.5)}, rotate=0]
               \draw[->,thick, mydarkgray] (0,0,0) -- (1,0,0) node[right] {$y_2$}; 
             \draw[->,thick, mydarkgray] (0,0,0) -- (0,1,0) node[above] {$z$}; 
              \draw[->,thick, mydarkgray] (0,0,0) -- (0,0,1.5) node[below left] {$y_1$}; 
              \end{scope}
\end{tikzpicture}
\caption{A cross-section of a 3D parallel-to-point system with two reflectors.}
\label{fig: 3D optical system}
\end{figure}
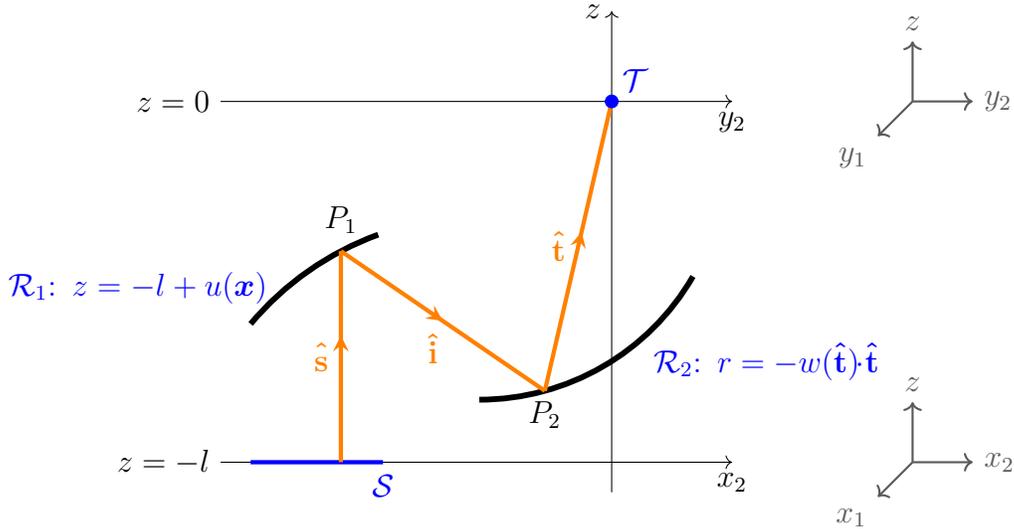
\noindent We present a mathematical model for the design of a double-reflector imaging system with a parallel source and a point target. A brief outline is as follows. First, the principles of geometrical optics are used to derive a relation of the form
 \begin{equation}\label{eq: 3D cost fxn}
 	u_1(\boldsymbol{x})+u_2(\boldsymbol{y})=c(\boldsymbol{x},\boldsymbol{y}),
 \end{equation}
 where the source and target are parameterized by variables $\boldsymbol{x}$ and $\boldsymbol{y}$, respectively, the functions $u_1$ and $u_2$ are related to the shapes and locations of the surfaces, and $c$ is the so-called cost function in the framework of optimal mass transport. Next, the law of conservation of energy, combined with a linear optical map in phase space, implies that the ratio of the energy distributions at the source and target is constant. Finally, the optical map is utilized to calculate the reflectors using optimal transport theory. 
\subsection{Cost function formulation}
Consider a 3D optical system consisting of two reflectors $\mathcal{R}_1,$ $\mathcal{R}_2$ and a set of parallel rays forming a point image in the target plane $z=0$ (see Fig.~\ref{fig: 3D optical system}). A parallel source located at $z=-l$ emits rays with unit direction vector $\boldsymbol{\hat{s}}=(s_1,s_2,s_3)^{\rm{T}}=(0,0,1)^{\rm{T}}.$ In the sequel, the symbol $(\,\boldsymbol{\hat{}}\,)$ is used to denote unit vectors. We choose the point target to be located at the origin $\mathcal{O}$ of the target coordinate system. The rays hit the target plane with unit direction vectors $\boldsymbol{\hat{t}}=(t_1,t_2,t_3)^{\rm{T}}.$ The first reflector, $\mathcal{R}_1: z=-l+u(\boldsymbol{x})$, is defined by its perpendicular distance $u(\boldsymbol{x})$ from the source plane. The second reflector, $\mathcal{R}_2: \boldsymbol{r}(\boldsymbol{\hat{t}})=-w(\boldsymbol{\hat{{t}}})\boldsymbol{\hat{t}}$, is defined by the radial distance  $w(\boldsymbol{\hat{{t}}})$ from the target. The variables $\boldsymbol{x}$ and $\boldsymbol{\hat{t}}$ denote the position of a ray at the source and the direction to the target, respectively. With $P_1$ and $P_2$, we denote the points where a ray hits the first and second reflectors, respectively. The distance between those two points is denoted by $d$. The positions of the ray at the source and target are given by $\boldsymbol{q}_{\rm{s}}=\boldsymbol{x}$ and $\boldsymbol{q}_{\rm{t}}=\boldsymbol{0},$ respectively. The momentum coordinates, i.e, the projections of the direction vectors on the source and target planes are $\boldsymbol{p}_{\rm{s}}=\boldsymbol{0}$ and $\boldsymbol{p}_{\rm{t}},$ respectively. 

In terms of Hamiltonian characteristics, the optical path length (OPL) is equal to the point characteristic $V$, i.e.,
\begin{equation}
	\label{eq:point ch}
	V(\boldsymbol{q}_{\rm{s}},\boldsymbol{q}_{\rm{t}})=u(\boldsymbol{x})+d+w(\boldsymbol{\hat{t}}).
\end{equation}
Since we work with position at the source and the direction to the target, we consider the mixed characteristic of first kind $W=W(\boldsymbol{q}_{\rm{s}},\boldsymbol{p}_{\rm{t}})$ \cite[Ch.~2]{lotte}, given by
\begin{equation}\label{eq: mixed ch first kind}
	W(\boldsymbol{q}_{\rm{s}},\boldsymbol{p}_{\rm{t}})=V(\boldsymbol{q}_{\rm{s}},\boldsymbol{q}_{\rm{t}})-\boldsymbol{q}_{\rm{t}}\boldsymbol{\cdot}\boldsymbol{p}_{\rm{t}}.
\end{equation}
Since $\boldsymbol{q}_{\rm{t}}=\boldsymbol{0}$, the mixed characteristic is equal to the OPL. The following relations hold for the mixed characteristic
\begin{equation}\label{eq: relns mixed ch}
	-\frac{\partial W}{\partial \boldsymbol{q}_{\rm{s}}}=\boldsymbol{p}_{\rm{s}}=\boldsymbol{0}, \quad 	-\frac{\partial W}{\partial \boldsymbol{p}_{\rm{t}}}=\boldsymbol{q}_{\rm{t}}=\boldsymbol{0}.
\end{equation}
This shows that $W$ is independent of the position on the source and the direction to the target. Thus, the OPL is a constant and $V=W$. We omit the dependence of $u$  and $w$ on $\boldsymbol{x}$ and $\boldsymbol{\hat{t}}$ for now, and Eq.~(\ref{eq:point ch}) implies
\begin{equation}\label{eq: d from V}
	d^2=(V-u-w)^2.
\end{equation}
The distance $d$ between the points $P_1$ and $P_2$ is given by  
\begin{equation}\label{eq: dist b/w points}
	d^2=(x_1+wt_1)^2+(x_2+wt_2)^2+(-l+u+wt_3)^2.
\end{equation}
We eliminate $d$ from Eqs.~(\ref{eq: d from V})-(\ref{eq: dist b/w points}) to obtain
\begin{equation}\label{eq: eliminate d}
    |\boldsymbol{x}|^2-(V-l)(V+l)+2u(V-l)+2w(\boldsymbol{x}\cdot\boldsymbol{p}_{\rm{t}})+2Vw-2lwt_3-2uw(1-t_3)=0,
\end{equation}
where $\boldsymbol{p}_{\rm{t}}=(t_1,t_2)^{\rm{T}}$, since the momentum vector $\vec{p}_{\rm{t}}$ is equal to the reflected direction $\boldsymbol{\hat{t}}$. We want to separate $u$ and $w$ to obtain an equation of the form of Eq.~(\ref{eq: 3D cost fxn}).  We introduce the reduced OPL denoted by $\beta=V-l>0$. Since $w>0$, we can divide Eq.~(\ref{eq: eliminate d}) by $2\beta w$. Then, Eq.~(\ref{eq: eliminate d}) is expressed as
\begin{equation}\label{eq: int reln by eliminating d}
	\frac{u}{w}+\frac{1}{2\beta w}\left(\left|\boldsymbol{x}\right|^2-\beta(V+l)\right)-\frac{u}{\beta}\left(1-t_3\right)=-\frac{1}{\beta}\left(\boldsymbol{x}\cdot\boldsymbol{p}_{\rm{t}}-lt_3+V\right).
\end{equation}
Eq.~(\ref{eq: int reln by eliminating d}) is factorized by adding the term $-(1-t_3)(\left|\boldsymbol{x}\right|^2-\beta(V+l))/2\beta^2$ on both sides of the equation
\begin{equation}\label{eq: factorize}
\eqalign{	 \left(u+\frac{1}{2\beta}\left(\left|\boldsymbol{x}\right|^2-\beta(V+l)\right)\right)\left(\frac{1}{w}-\frac{1-t_3}{\beta}\right)=\cr
\qquad  -\frac{1}{\beta}\Big(\boldsymbol{x}\cdot\boldsymbol{p}_{\rm{t}}-lt_3+V\Big)-\frac{1-t_3}{2\beta^2}\left(\left|\boldsymbol{x}\right|^2-\beta(V+l)\right).}
\end{equation}
We introduce the following nomenclature for Eq.~(\ref{eq: factorize}) 
\begin{equation}\label{eq: kappa}
	\kappa_1(\boldsymbol{x})\kappa_2(\boldsymbol{\hat{t}})=A(\boldsymbol{x},\boldsymbol{\hat{t}}),
\end{equation}

We investigate whether $\kappa_1(\boldsymbol{x})$, $\kappa_2(\boldsymbol{\hat{t}})$ and $A(\boldsymbol{x},\boldsymbol{\hat{t}})$ are positive or negative because we want to take the logarithm of these arguments to obtain an equation as in Eq.~(\ref{eq: 3D cost fxn}). From Fig.~\ref{fig: 3D optical system} we conclude that
$l=(u\boldsymbol{\hat{s}}+d\boldsymbol{\hat{i}}+w\boldsymbol{\hat{t}})\cdot\boldsymbol{\hat{e}}_3=u+di_3+wt_3.$ The reduced OPL $\beta=V-l$ can be expressed as
\begin{equation}
    \label{eq: beta n w reln}
	\beta=d(1-i_3)+w(1-t_3).
\end{equation}
Since $i_3$ and $t_3$ are the third components of the unit vectors $\boldsymbol{\hat{i}}$ and $\boldsymbol{\hat{t}}$, respectively, it follows that $\left|i_3\right|\leq1$ and $\left|t_3\right|\leq1$. This leads to $1-i_3\geq0$, where the equality does not hold, as otherwise there is no reflection on the first surface. Also, $1-t_3\geq0$. From Eqs.~(\ref{eq: factorize})-(\ref{eq: beta n w reln}) we obtain
\begin{equation}\label{eq: sign kappa2}
		\kappa_2(\boldsymbol{\hat{t}})=\frac{1}{w}-\frac{1-t_3}{d(1-i_3)+w(1-t_3)}=\frac{1}{w}-\frac{1}{d\left(\frac{1-i_3}{1-t_3}\right)+w}>0.
\end{equation}

To determine the sign of  $A(\boldsymbol{x},\boldsymbol{\hat{t}})$, we simplify the expression in Eq.~(\ref{eq: factorize}) by parameterizing the unit direction vector $\boldsymbol{\hat{t}}$ by its stereographic projection from the south pole \cite[p.~60-62]{lotte}. The stereographic projection $\boldsymbol{y}$ and the corresponding inverse projection $\boldsymbol{\hat{t}}$ are given by
\begin{equation}\label{eq: st proj 3D}
\boldsymbol{y}=\frac{1}{1+t_3}\left[\begin{array}{cc}t_1\\t_2\end{array}\right],\quad \boldsymbol{\hat{t}}=\frac{1}{1+\left|\boldsymbol{y}\right|^2}\left[\begin{array}{cc} 2y_1\\2y_2\\1-\left|\boldsymbol{y}\right|^2\end{array}\right].
\end{equation}
We recall that $\boldsymbol{\vec{p}}_{\rm{t}}=\boldsymbol{\hat{t}}$. Using Eq.~(\ref{eq: 3D pz}), we have $t_3=\pm \sqrt{1-|\boldsymbol{p}_{\rm{t}}|^2}$. From Fig.~\ref{fig: 3D optical system}, we observe that $\boldsymbol{\hat{t}}$ points upwards, which means that the positive sign holds for $t_3$. Then, Eq.~(\ref{eq: st proj 3D}) leads to
\begin{equation}\label{eq: 3D pt to y}
 \boldsymbol{y}=\frac{\boldsymbol{p}_{\rm{t}}}{1+\sqrt{1-|\boldsymbol{p}_{\rm{t}}}|^2}.
\end{equation}
The relations in Eq.~(\ref{eq: st proj 3D}) are used to transform $A(\boldsymbol{x},\boldsymbol{\hat{t}})$ to $A(\boldsymbol{x},\boldsymbol{y})$
\begin{equation}
 A(\boldsymbol{x},\boldsymbol{y})=-\frac{1}{1+\left|\boldsymbol{y}\right|^2}\left(1+\frac{2\,\boldsymbol{x}\boldsymbol{\cdot}\boldsymbol{y}}{\beta}+\frac{\left|\boldsymbol{x}\right|^2\left|\boldsymbol{y}\right|^2}{\beta^2}\right).
    \label{eq: A(x,t) to A(x,y)}
\end{equation}
From the Cauchy-Schwarz inequality, we know that $|\boldsymbol{x}\boldsymbol{\cdot}\boldsymbol{y}|\leq |\boldsymbol{x}|\,|\boldsymbol{y}|$.  Then, using Eq.~(\ref{eq: A(x,t) to A(x,y)}) we determine the sign of $A(\boldsymbol{x},\boldsymbol{y})$

\begin{equation}
	\hspace{-2.5cm}A(\boldsymbol{x},\boldsymbol{y})
    \leq-\frac{1}{1+\left|\boldsymbol{y}\right|^2}\left(1-\frac{2\,\left|\boldsymbol{x}\right|\left|\boldsymbol{y}\right|}{\beta}+\frac{\left|\boldsymbol{x}\right|^2\left|\boldsymbol{y}\right|^2}{\beta^2}\right)
    =-\frac{1}{1+\left|\boldsymbol{y}\right|^2}\left(1-\frac{\left|\boldsymbol{x}\right|\left|\boldsymbol{y}\right|}{\beta}\right)^2<0.\label{eq: sign A(x,y)}
\end{equation}
The expression from $\kappa_2(\boldsymbol{\hat{t}})$ can be expressed as $\kappa_2(\boldsymbol{y})$ by substituting Eq.~(\ref{eq: st proj 3D}) in Eq.~(\ref{eq: factorize})
\begin{equation}\label{eq: kappa2(y)}
    \kappa_2(\boldsymbol{y})=\frac{1}{w}-\frac{2\left|\boldsymbol{y}\right|^2}{\beta(1+\left|\boldsymbol{y}\right|^2)}.
\end{equation}
Since $\kappa_2(\boldsymbol{\hat{t}})>0$, the expression $\kappa_2(\boldsymbol{y})$ in Eq.~(\ref{eq: kappa2(y)}) is also positive.

We transformed $\boldsymbol{\hat{t}}$ to the stereographic projection $\boldsymbol{y}$, the expressions $\kappa_2(\boldsymbol{\hat{t}}(\boldsymbol{y}))$ and $A(\boldsymbol{x},\boldsymbol{\hat{t}})$ can be rewritten as $\kappa_2(\boldsymbol{y})$ and $A(\boldsymbol{x},\boldsymbol{y})$. Thus, from Eq.~(\ref{eq: kappa}), we have $\kappa_1(\boldsymbol{x})\kappa_2(\boldsymbol{y})=A(\boldsymbol{x},\boldsymbol{y})$. From Eqs.~(\ref{eq: sign kappa2}) and (\ref{eq: sign A(x,y)}), we know that $\kappa_2((\boldsymbol{y})>0$ and $A(\boldsymbol{x},\boldsymbol{y})<0$. Consequently, $\kappa_1(\boldsymbol{x})<0$. We multiply the negative expressions by $-1$, which leads to $-\kappa_1(\boldsymbol{x})\kappa_2(\boldsymbol{y})=-A(\boldsymbol{x},\boldsymbol{y}).$ We take the logarithm on both sides of preceeding relation to obtain the desired form in Eq.~(\ref{eq: 3D cost fxn}) and using Eqs.~(\ref{eq: factorize}), (\ref{eq: A(x,t) to A(x,y)}) and (\ref{eq: kappa2(y)}), we obtain the following 
\numparts\label{eq: 3D expressions u1,u2,c}
\begin{equation}
	u_1(\boldsymbol{x})=\log\left(-u-\frac{\left|\boldsymbol{x}\right|^2}{2\beta}+\frac{V+l}{2}\right),\label{eq: 3D u1}\end{equation}
    \begin{equation}
	u_2(\boldsymbol{y})=\log\left(\frac{1}{w}-\frac{2\left|\boldsymbol{y}\right|^2}{\beta(1+\left|\boldsymbol{y}\right|^2)}\right),\label{eq: 3D u2}
    \end{equation}
    \begin{equation}
	c(\boldsymbol{x},\boldsymbol{y})=\log\left(\frac{1}{1+\left|\boldsymbol{y}\right|^2}\left(1+\frac{2\,\boldsymbol{x}\boldsymbol{\cdot}\boldsymbol{y}}{\beta}+\frac{\left|\boldsymbol{x}\right|^2\left|\boldsymbol{y}\right|^2}{\beta^2}\right)\right).\label{eq: 3D c}
\end{equation}
\endnumparts

\subsection{Energy conservation}
 Let $f(\boldsymbol{q}_{\mathrm{s}})$ be the exitance at the source and $g(\boldsymbol{p}_{\mathrm{t}})$ be the intensity at the target. Let $\boldsymbol{\tilde{m}}$ denote the phase-space optical map that connects the source coordinate $\boldsymbol{q}_{\rm{s}}\in Q_{\mathrm{s}}$ to the target momentum coordinate $\boldsymbol{p}_{\rm{t}}\in P_{\mathrm{t}},$ where $\boldsymbol{\tilde{m}}(\boldsymbol{q}_{\rm{s}})=M_2(\boldsymbol{q}_{\rm{s}},\boldsymbol{0})$; see Eq.~(\ref{eq: 3D phase-space optical map general}). The law of conservation of energy reads 
\begin{equation}\label{eq: 3D energy cons phase space 2}
	 \int_{\tilde{\mathcal{A}}} f(\boldsymbol{q}_{\mathrm{s}})\,\mathrm{d}\boldsymbol{q}_{\mathrm{s}}=	\int_{\boldsymbol{\tilde{m}}(\tilde{\mathcal{A}})}  g(\boldsymbol{p}_{\mathrm{t}})\,\mathrm{d}\boldsymbol{p}_{\mathrm{t}},
\end{equation}
for any subset $\tilde{\mathcal{A}}\subseteq Q_{\mathrm{s}}$ and image set $\boldsymbol{\tilde{m}}(\tilde{\mathcal{A}})\subseteq P_{\mathrm{t}}$. In any optical system, the total flux of the source equals that of the target, implying that energy must be conserved globally. Eq.~(\ref{eq: 3D energy cons phase space 2}) holds for all subsets $\tilde{\mathcal{A}}\subseteq Q_{\mathrm{s}}$ and $\boldsymbol{\tilde{m}}(\tilde{\mathcal{A}})\subseteq P_{\mathrm{t}}$, and when $\tilde{\mathcal{A}}=Q_{\mathrm{s}}$ and $\boldsymbol{\tilde{m}}(\tilde{\mathcal{A}})= P_{\mathrm{t}}$, we have global energy conservation. Substituting the mapping $\boldsymbol{\tilde{m}}(\boldsymbol{q}_{\mathrm{s}})=\boldsymbol{p}_{\mathrm{t}}$ in Eq.~(\ref{eq: 3D energy cons phase space 2}) results in
\begin{equation}\label{eq: 3D phase-space PDE}
    \det(\mathrm{D}\boldsymbol{\tilde{m}}(\boldsymbol{q}_{\rm{s}}))=\pm\frac{f(\boldsymbol{q}_{\rm{s}})}{g(\boldsymbol{\tilde{m}}(\boldsymbol{q}_{\rm{s}}))},
\end{equation}
subject to the transport boundary condition $\boldsymbol{\tilde{m}}(\partial Q_{\rm{s}})=\partial P_{\rm{t}}$, which ensures that the edge-ray principle is satisfied \cite{TBC}, implying the boundary of the source domain is mapped to the boundary of the target domain.

For the optical map $\boldsymbol{\tilde{m}}: \boldsymbol{q}_{\rm{s}}\mapsto \boldsymbol{p}_{\rm{t}}$, we know that $\det\left(\mathrm{D}\boldsymbol{\tilde{m}}(\boldsymbol{q}_{\rm{s}})\right)=\det\left(\partial \boldsymbol{p}_{\rm{t}}/\partial \boldsymbol{q}_{\rm{s}}\right)$. Since we aim to design an imaging system, we impose the phase-space optical map that ensures ideal image formation. From Eq.~(\ref{eq: 3D linear matrix for imaging}), it follows that for a parallel-to-point optical system, the phase-space optical map is given by $\boldsymbol{\tilde{m}}(\boldsymbol{q}_{\rm{s}})=\boldsymbol{p}_{\rm{t}}=\mathcal{C}\boldsymbol{q}_{\rm{s}}$, since $\boldsymbol{p}_{\rm{s}}=\boldsymbol{0}$ and $\boldsymbol{q}_{\rm{t}}=\boldsymbol{0}$. Then, Eq.~(\ref{eq: corresponding ABCD}) leads to $\det(\mathrm{D}\boldsymbol{\tilde{m}}(\boldsymbol{q}_{\rm{s}}))=\det(\mathcal{C})$. Therefore, from Eq.~(\ref{eq: 3D phase-space PDE}), we conclude that the ratio of energy distributions at the source and target of the optical system is equal to a constant, i.e., $\det(\mathcal{C})$. 

\subsection{Optical map for imaging}\label{sec: optical map for imaging}
 This paper focuses on designing a rotationally symmetric imaging system. In this section, our goal is to determine the structure of the associated matrix $\mathcal{C}$, and subsequently, the corresponding optical map $\boldsymbol{\tilde{m}}$. Since $\mathcal{C}$ governs the optical map $\boldsymbol{\tilde{m}}$, we assume $\mathcal{C}\in \mathbb{R}^{2\times2}$. Let $\mathcal{Q}$ be a $2\times2$ matrix describing rotation around the optical axis. We first show that the matrix $\mathcal{C}$ commutes with the rotation matrix. Let $\breve{\boldsymbol{q}}_{\rm{s}}=\mathcal{Q}\boldsymbol{q}_{\rm{s}}$ denote the rotated source and $\breve{\boldsymbol{p}}_{\rm{t}}$ be the corresponding target, then the following holds
\begin{equation}\label{eq: 3D rotated map}
    \breve{\boldsymbol{p}}_{\rm{t}}=\mathcal{C}\breve{\boldsymbol{q}}_{\rm{s}}.
\end{equation}
Since the system is rotationally symmetric, the phase-space optical map is invariant under rotation, implying that $\breve{\boldsymbol{p}}_{\rm{t}}=\mathcal{Q}\boldsymbol{p}_{\rm{t}}.$ Eq.~(\ref{eq: 3D rotated map}) can be written as
\begin{equation}\label{eq: 3D rotated map 2}
 \mathcal{ Q}\boldsymbol{p}_{\rm{t}}=\mathcal{C}\mathcal{Q}\boldsymbol{q}_{\rm{s}}.
\end{equation}
Since $\boldsymbol{p}_{\rm{t}}=\mathcal{C}\boldsymbol{q}_{\rm{s}}$, Eq.~(\ref{eq: 3D rotated map 2}) results in $\mathcal{Q}\mathcal{C}\boldsymbol{q}_{\rm{s}}=\mathcal{C}\mathcal{Q}\boldsymbol{q_{\rm{s}}}$, implying $\mathcal{C}\mathcal{Q}=\mathcal{Q}\mathcal{C}.$ 

The matrix $\mathcal{C}$ can be represented uniquely as the sum of a symmetric matrix $\mathcal{H}$ and an antisymmetric matrix $\mathcal{{H}}^{a}$
\begin{equation}\label{eq: C sym and antisym}
\mathcal{C}=\mathcal{H}+\mathcal{H}^a,
\end{equation}
Since $\mathcal{C}$ commutes with all rotation matrices, $\mathcal{C}=\mathcal{Q}^{\rm{T}}\mathcal{C}\mathcal{Q}$. Then, $\mathcal{C}$ can be expressed as
 \begin{equation}\label{eq: C sym and antisym with Q}
\mathcal{C}=\mathcal{Q}^{\rm{T}}\mathcal{H}\mathcal{Q}+\mathcal{Q}^{\rm{T}}\mathcal{{H}}^a\mathcal{Q},
\end{equation} 
For the symmetric matrix $\mathcal{H}$, we know that $\mathcal{H}^{\rm{T}}=\mathcal{H}$, implying $(\mathcal{Q}^{\rm{T}}\mathcal{H}\mathcal{Q})^{\rm{T}}=\mathcal{Q}^{\rm{T}}\mathcal{H}\mathcal{Q}$. Also, $(\mathcal{{H}}^a)^{\rm{T}}=-\mathcal{{H}}^a$ holds for the antisymmetric matrix $\mathcal{{H}}^a$, resulting in $(\mathcal{Q}^{\rm{T}}\mathcal{{H}}^a\mathcal{Q})^{\rm{T}}=-\mathcal{Q}^{\rm{T}}\mathcal{{H}}^a\mathcal{Q}$. This means that $\mathcal{Q}^{\rm{T}}\mathcal{H}\mathcal{Q}$ and $\mathcal{Q}^{\rm{T}}\mathcal{{H}}^a\mathcal{Q}$ are symmetric and antisymmetric matrices, respectively. Since Eq.~(\ref{eq: C sym and antisym}) holds true for unique $\mathcal{H}$ and $\mathcal{{H}}^a$, from Eqs.~(\ref{eq: C sym and antisym}) and (\ref{eq: C sym and antisym with Q}) we conclude that 
\begin{equation}\label{eq: h and bar h unique}
\mathcal{H}=\mathcal{Q}^{\rm{T}}\mathcal{H}\mathcal{Q}, \quad 
\mathcal{{H}}^a=\mathcal{Q}^{\rm{T}}\mathcal{{H}}^a\mathcal{Q}.
\end{equation}

Let $\boldsymbol{v}\neq\boldsymbol{0}\in\mathbb{R}^2$ be an arbitrary eigenvector of the matrix $\mathcal{H}$, with an eigenvalue $k_1\in \mathbb{R}$, then
\begin{equation}\label{eq: eigenvalue}
\mathcal{H}\boldsymbol{v}=k_1\boldsymbol{v}.
\end{equation}
From Eqs.~(\ref{eq: h and bar h unique}) and (\ref{eq: eigenvalue}), we have the following 
\begin{equation}
 \mathcal{H}(\mathcal{Q}\boldsymbol{v})=\mathcal{H}\mathcal{Q}\boldsymbol{v}
    =\mathcal{Q}\mathcal{H}\boldsymbol{v}
    =k_1\mathcal{Q}\boldsymbol{v}.
\end{equation}
Thus, $\mathcal{Q}\boldsymbol{v}$ is also an eigenvector of $\mathcal{H}.$ $\mathcal{Q}\boldsymbol{v}$ can be any vector in $\mathbb{R}^2$, implying that the matrix $\mathcal{H}$ has the same eigenvalue $k_1\in \mathbb{R}$ for every vector in $\mathbb{R}^2.$ Therefore, the geometric multiplicity of $k_1$ is 2. Let ${h_{ij}}$, for $i,j=1,2,$ denote the components of $\mathcal{H}$. Since $\mathcal{H}$ is symmetric, $h_{12}=h_{21}$. The characterstic polynomial $\tilde{P}(\lambda)=\det({\mathcal{H}-\lambda I_{2\times2}})$ is
\begin{equation}\label{eq: ch poly}
\lambda^2-(h_{11}+h_{22})\lambda+h_{11}h_{22}-h_{12}^2=0,
\end{equation}
where $\lambda$ denotes an eigenvalue. Since $k_1$ is an eigenvalue of $\mathcal{H}$, it is a root of the characterstic polynomial $\tilde{P}(\lambda)$, i.e., $\det(\mathcal{H}-k_1I_{2\times2})=0.$ Given that $k_1$ is the only eigenvalue of $\mathcal{H}$, its algebraic multiplicity is $2$. The eigenvalue $k_1$ is a double-root of $\tilde{P}(\lambda)$, implying that the discriminant of Eq.~(\ref{eq: ch poly}) is zero, i.e,  $(h_{11}-h_{22})^2+4h_{12}^2=0$. This is possible when $h_{12}=0$ and $h_{11}=h_{22}$. Therefore, we conclude that 
\begin{equation}\label{eq: sym}
\mathcal{H}=k_1I_{2\times2}.
\end{equation}
The geometric multiplicity of any eigenvalue is always less than or equal to its algebraic multiplicity. For $\mathcal{H}$, both geometric and algebraic multiplicities of $k_1$ are equal to 2, and we conclude that a symmetric matrix $\mathcal{H}$ that commutes with all rotation matrices is a scalar multiple of an identity matrix.

The antisymmetric matrix $\mathcal{{H}}^a$ is given by 
\begin{equation}\label{eq: antisym}
\mathcal{{H}}^a=k_2J_{2\times2}, \quad J_{2\times2}=\left[\begin{array}{cc} 0&1\\-1&0 \end{array}\right].
\end{equation}

From Eqs.~(\ref{eq: sym}) and (\ref{eq: antisym}), we know that $\mathcal{C}=k_1I_{2\times2}+k_2J_{2\times2}$. $\mathcal{C}$ can be expressed as
\begin{equation}\label{eq: intermediate C}
\mathcal{C}=|\boldsymbol{k}|\left[\begin{array}{rc} \frac{k_1}{|\boldsymbol{k}|}&\frac{k_2}{|\boldsymbol{k}|}\\ -\frac{k_2}{|\boldsymbol{k}|}&\frac{k_1}{|\boldsymbol{k}|} \end{array}\right].
\end{equation}
where $\boldsymbol{k}=(k_1,k_2)^{\rm{T}}$ and  $|\boldsymbol{k}|=\sqrt{k_1^2+k_2^2}$.
Let us denote a rotation matrix describing rotation about an angle $\varphi$ with $\mathcal{Q}(\varphi)$. Then, Eq.~(\ref{eq: intermediate C}) leads to $\mathcal{C}=|\boldsymbol{k}|\,\mathcal{Q}(-\varphi)$, where $\cos\varphi=k_1/|\boldsymbol{k}|$ and $\sin\varphi=k_2/|\boldsymbol{k}|$. We introduce the notation $|\boldsymbol{k}|=k$ and express the phase optical map $\boldsymbol{p}_{\mathrm{t}}=\boldsymbol{\tilde{m}}(\boldsymbol{q}_{\mathrm{s}})$ as
\begin{equation}\label{eq: 3D phase space optical map intermediate}
\boldsymbol{\tilde{m}}(\boldsymbol{q}_{\mathrm{s}})=\mathcal{C}\boldsymbol{q}_{\mathrm{s}}=k\,\mathcal{Q}(-\varphi)\,\boldsymbol{q}_{\mathrm{s}}.
\end{equation}

The optimal transport formulation given by Eq.~(\ref{eq: 3D cost fxn}) enables us to design reflectors that connect the source coordinates $\boldsymbol{x}=(x_1,x_2)^{\rm{T}}$ to the stereographic target coordinates $\boldsymbol{y}=(y_1,y_2)^{\rm{T}}$. We know that $\boldsymbol{q}_{\mathrm{s}}=\boldsymbol{x}$, and we transform the momentum $\boldsymbol{p}_{\mathrm{t}}$ to the stereographic projection $\boldsymbol{y}$ to determine the optical map $\boldsymbol{y}=\boldsymbol{m}(\boldsymbol{x})$ that maps the source domain $\mathcal{S}$ to the stereographic target domain $\mathcal{T}$.  We substitute $\boldsymbol{q}_{\mathrm{s}}=\boldsymbol{x}$ in the phase-space optical map $\boldsymbol{\tilde{m}}$ in Eq.~(\ref{eq: 3D phase space optical map intermediate}) and subsequently use Eq.~(\ref{eq: 3D pt to y}) to determine the optical map $\boldsymbol{y}=\boldsymbol{m}(\boldsymbol{x})$  
\begin{equation}\label{eq: 3D optical map in terms of x}
 \boldsymbol{m}(\boldsymbol{x})=\frac{k\,\mathcal{Q}(-\varphi)\,\boldsymbol{x}}{1+\sqrt{1-k^2\left|\boldsymbol{x}\right|^2}}, \quad k|\boldsymbol{x}|\leq 1.
\end{equation}

 We consider rays emitted from a circular source domain $\mathcal{S}$. Let $|\boldsymbol{x}|\leq r$, for $r\in \mathbb{R}^{+}$, then from Eq.~(\ref{eq: 3D optical map in terms of x}), we obtain $ |\boldsymbol{m}(\boldsymbol{x})|\leq kr/1+\sqrt{1-(kr)^2}.$ Therefore, the stereographic target domain $\mathcal{T}$ will be a circle. So, we switch to polar coordinate systems. The polar source and stereographic target coordinates are written as $\boldsymbol{\zeta}=(r,\vartheta)$ and $\boldsymbol{\eta}=(R,\Theta)$ and the transformations $\boldsymbol{x}=\boldsymbol{x}(\boldsymbol{\zeta})$ and $\boldsymbol{y}=\boldsymbol{y}(\boldsymbol{\eta})$ are given by

    \begin{equation}
    \eqalign{
x_1=r\cos\vartheta, \qquad &x_2=r\sin\vartheta,\\
 y_1=R\cos\Theta, \quad &y_2=R\sin\Theta.}
\end{equation}
We define $\omega$ and $\rho$ to be the source and stereographic target domains in polar coordinates, so $\omega=\boldsymbol{\zeta}(\mathcal{S})$ and $\rho=\boldsymbol{\eta}(\mathcal{T})$. Let $\boldsymbol{m}^\star(\boldsymbol{\zeta})=\boldsymbol{m}(\boldsymbol{x}(\boldsymbol{\zeta}))$ denote the optical map in polar coordinates. For a rotationally symmetric system, $\boldsymbol{x}=r\boldsymbol{\hat{e}}_r$, where $\boldsymbol{\hat{e}}_r$ denotes the radial unit basis vector. Then Eq.~(\ref{eq: 3D optical map in terms of x}) results in the so-called polar optical map $\boldsymbol{y}=\boldsymbol{m}^\star(\boldsymbol{\zeta})$
\begin{equation}\label{eq: 3D optical map polar coordinates}
      \boldsymbol{m}^\star(\boldsymbol{\zeta})=\frac{k\,\mathcal{Q}(-\varphi)\,r\boldsymbol{\hat{e}}_r}{1+\sqrt{1-k^2r^2}},
\end{equation}
In the remainder of this paper, we will omit the asterisk because we will only work with the optical map in polar coordinates.

\subsection{Freeform reflectors}\label{sec: freeform reflectors}

We define $\tilde{u}_1(\boldsymbol{\zeta})=u_1(\boldsymbol{x}(\boldsymbol{\zeta}))$, and $\tilde{c}(\boldsymbol{\zeta},\boldsymbol{y})=c(\boldsymbol{x}(\boldsymbol{\zeta}),\boldsymbol{y})$. Eq.~(\ref{eq: 3D cost fxn}) changes to 
\begin{equation}\label{eq: 3D cost fxn polar}
    \tilde{u}_1\boldsymbol{\zeta})+u_2(\boldsymbol{y})=\tilde{c}(\boldsymbol{\zeta},\boldsymbol{y}).
\end{equation}


Eq.~(\ref{eq: 3D cost fxn polar}) has many possible solutions, but we restrict ourselves to the following solution pairs \cite{yadav} 
\numparts 
    \begin{equation}
 \hspace{-2.5cm} \mathrm{c-convex:}	\quad	\tilde{u}_1(\boldsymbol{\zeta})=\mathop{\mathrm{max}}\limits_{\boldsymbol{y}\in\mathcal{T}}\Big(\tilde{c}(\boldsymbol{\zeta},\boldsymbol{y})-u_2(\boldsymbol{y})\Big), \quad	u_2(\boldsymbol{y})=\mathop{\mathrm{max}}\limits_{\boldsymbol{\zeta}\in\omega}\Big(\tilde{c}(\boldsymbol{\zeta},\boldsymbol{y})-\tilde{u}_1(\boldsymbol{\zeta})\Big),
  	   \label{eq: convexsoln}\end{equation}
       \begin{equation}
 \hspace{-2.5cm}        \mathrm{c-concave:}	\quad	\tilde{u}_1(\boldsymbol{\zeta})=\mathop{\mathrm{min}}\limits_{\boldsymbol{y}\in\mathcal{T}}\Big(\tilde{c}(\boldsymbol{\zeta},\boldsymbol{y})-u_2(\boldsymbol{y})\Big), \quad	u_2(\boldsymbol{y})=\mathop{\mathrm{min}}\limits_{\boldsymbol{\zeta}\in\omega}\Big(\tilde{c}(\boldsymbol{\zeta},\boldsymbol{y})-\tilde{u}_1(\boldsymbol{\zeta})\Big).
  	   \label{eq: concavesoln}
  	\end{equation}
    \endnumparts
  	Eqs.~(\ref{eq: convexsoln})-(\ref{eq: concavesoln}) require $\boldsymbol{\zeta}$ to be a stationary point of $\tilde{c}(\cdot,\boldsymbol{y})-\tilde{u}_1$. As a result, a necessary condition for the existence of the solution is
     \begin{equation}\label{eq: 3D necessary condition polar}
     \nabla_{\boldsymbol{\zeta}}\tilde{c}(\boldsymbol{\zeta},\boldsymbol{y})-\nabla_{\boldsymbol{\zeta}} \tilde{u}_1(\boldsymbol{\zeta})=\boldsymbol{0}.
      \end{equation} 
      In Eq.~(\ref{eq: 3D necessary condition polar}), the gradient with respect to $\boldsymbol{\zeta}$ is given by
\begin{equation}\label{eq: diff in polar}
    \nabla_{\boldsymbol{\zeta}}=\frac{\partial }{\partial r}\boldsymbol{\hat{e}}_r+\frac{1}{r}\frac{\partial }{\partial \vartheta}\boldsymbol{\hat{e}}_{\vartheta},
\end{equation}
where $\boldsymbol{\hat{e}}_r$ and $\boldsymbol{\hat{e}}_{\vartheta}$ denote the radial and angular unit basis vectors, respectively. Since $\boldsymbol{\hat{e}}_r$ and $\boldsymbol{\hat{e}}_{\vartheta}$ are linearly independent vectors, Eq.~(\ref{eq: diff in polar}) implies that Eq.~(\ref{eq: 3D necessary condition polar}) holds only when the following first-derivative terms are zero
\begin{equation}\label{eq: first deriavtives 0}
\frac{\partial }{\partial r}\big(\tilde{c}(\boldsymbol{\zeta},\boldsymbol{y})-\tilde{u}_1(\boldsymbol{\zeta})\big)=0, \quad \frac{\partial }{\partial \vartheta}\left(\tilde{c}(\boldsymbol{\zeta},\boldsymbol{y})-\tilde{u}_1(\boldsymbol{\zeta})\right)=0.
\end{equation}
Using Eq.~(\ref{eq: 3D c}), we determine $\tilde{c}(\boldsymbol{\zeta},\boldsymbol{y})$ and calculate the first-order partial derivatives
\numparts\label{eq: 3D partial derivatives}
\begin{equation}
\tilde{c}(\boldsymbol{\zeta},\boldsymbol{y})= \log\left(\frac{\beta^2+2 \beta(r\boldsymbol{\hat{e}_r})\cdot\boldsymbol{y}+r^2|\boldsymbol{y}|^2}{\beta^2(1+|\boldsymbol{y}|^2)}\right),\label{eq: polar c }\end{equation}
\begin{equation}
\quad\frac{\partial \tilde{c}} {\partial r}= \frac{2(\beta \, \boldsymbol{\hat{e}}_{r}\cdot\boldsymbol{y}+r|\boldsymbol{y}|^2)}{\beta^2+2\beta (r\boldsymbol{\hat{e}}_{r})\cdot\boldsymbol{y}+r^2|\boldsymbol{y}|^2},\label{eq: partial derivative of polar c wrt r}\end{equation}
\begin{equation}
\quad\frac{\partial \tilde{c}} {\partial \vartheta}=\frac{2\beta \, (r\boldsymbol{\hat{e}}_{\vartheta})\cdot\boldsymbol{y}}{\beta^2+2\beta (r\boldsymbol{\hat{e}}_{r})\cdot\boldsymbol{y}+r^2|\boldsymbol{y}|^2}. \label{eq: partial derivative of polar c wrt angle}
\end{equation}
\endnumparts
 We transform Eq.~(\ref{eq: 3D u1}) to polar coordinates to determine the expression $\tilde{u}_1(\boldsymbol{\zeta})$. It follows that for a rotationally symmetric system, $\partial \tilde{u}_1/\partial \vartheta=0$. Thus, Eq.~(\ref{eq: first deriavtives 0}) implies $\partial \tilde{c}/\partial \vartheta=0$. Then, Eq.~(\ref{eq: partial derivative of polar c wrt angle}) leads to the condition 
 \begin{equation}\label{eq: condition}
2\beta(r\boldsymbol{\hat{e}}_{\vartheta})\cdot\boldsymbol{y}=0.
 \end{equation}
 Next, we substitute the polar optical map $\boldsymbol{y}=\boldsymbol{m}(\boldsymbol{\zeta})$ given by Eq.~(\ref{eq: 3D optical map polar coordinates}) in Eq.~(\ref{eq: condition}). Since $\beta,r>0$, Eq.~(\ref{eq: condition}) holds when $ \boldsymbol{\hat{e}}_{\vartheta}\cdot \mathcal{Q}(-\varphi) \boldsymbol{\hat{e}}_{r}=0$, implying $\mathcal{Q}(-\varphi) \boldsymbol{\hat{e}}_{r}=\pm\boldsymbol{\hat{e}}_{r}.$ This is possible only when $\sin\varphi=0$, which means that $\varphi=n\pi$, where $n\in\mathbb{Z}$. Consequently, we have
 \begin{equation}\label{eq: C nature}
  \mathcal{C}=\pm kI_{2\times2}, \quad \det(\mathcal{C})=k^2.
 \end{equation}
It follows that only the positive sign holds for Eq.~(\ref{eq: 3D phase-space PDE}), resulting in 
\begin{equation}
    \det(\mathrm{D}\boldsymbol{\tilde{m}}(\boldsymbol{q}_{\rm{s}}))=\frac{f(\boldsymbol{q}_{\rm{s}})}{g(\boldsymbol{\tilde{m}}(\boldsymbol{q}_{\boldsymbol{s}}))}=k^2.
\end{equation}
Thus, for the inverse design of an imaging system, the ratio of energy distributions at the source and target must be constant, i.e., $f/g=k^2>0$. In Sec.~\ref{sec: inverse imaging}, we will show that the value of the constant $k^2$ is determined using energy conservation. Since Eq.~(\ref{eq: C nature}) holds for any $|k|=\sqrt{k^2}$, and we will only calculate $k^2$, choosing an appropriate sign for $k$ is essential. This choice will be discussed in Sec.~\ref{sec: inverse imaging}.

Next, we substitute the matrix $\mathcal{C}$ from Eq.~(\ref{eq: C nature}) in Eq.~(\ref{eq: 3D phase space optical map intermediate}) and obtain the phase-space optical map $\boldsymbol{\tilde{m}}$ of a parallel-to-point imaging system 
\begin{equation}\label{eq: final 3D linear phase space map}
       \boldsymbol{p}_{\rm{t}}=\boldsymbol{\tilde{m}}(\boldsymbol{q}_{\rm{s}})=\pm k\boldsymbol{q}_{\rm{s}}.
\end{equation}
Eq.~(\ref{eq: 3D optical map polar coordinates}) results in the polar optical map $\boldsymbol{m}: \boldsymbol{\zeta}\mapsto \boldsymbol{y}$
 \begin{equation}
    \boldsymbol{m}(\boldsymbol{\zeta})=\frac{\pm kr\boldsymbol{\hat{e}}_r}{1+\sqrt{1-k^2r^2}} .
  	   \label{eq: final 3D polar optical map}
 \end{equation}
Substitution of Eq.~(\ref{eq: final 3D polar optical map}) in Eq.~(\ref{eq: partial derivative of polar c wrt r}), combined with Eq.~(\ref{eq: first deriavtives 0}) leads to 
\begin{equation}\label{eq: 3D u1 ODE}
    \frac{\mathrm{d}\tilde{u}_1}{\mathrm{d}r}=\frac{\pm 2kr}{\pm kr^2+\beta(1+\sqrt{1-(kr)^2})}.
\end{equation}
We choose the value of $\tilde{u}_1$ at one point. This condition, together with  Eq.~(\ref{eq: 3D u1 ODE}), gives an initial value problem, which can be solved numerically to find a unique solution. In this paper, we choose the value of $\tilde{u}_1$ along the central ray and denote it by $u_0.$ This fixes the position of the vertex of the first reflector. Next, we obtain $u_2$ from Eq.~(\ref{eq: 3D cost fxn polar}).

We transform Eq.~(\ref{eq: 3D u2}) to the polar coordinates and determine the expression $\tilde{u}_2(\boldsymbol{\eta})=u_2(\boldsymbol{y}(\boldsymbol{\eta}))$. Consequently, we determine the shapes of the reflectors, given by
\numparts\label{eq: 3D reflector shapes}
   \begin{equation}
      u(r,\vartheta)=-\exp(\tilde{u}_1)-\frac{r^2}{2\beta}+\frac{V+l}{2},\label{eq: 3D first reflector}\end{equation}
        \begin{equation}
     w(R,\Theta)=\left(\exp(\tilde{u}_2)+\frac{2R^2}{\beta(1+R^2)}\right)^{-1}.\label{eq: 3D second reflector}
   \end{equation}
\endnumparts

\section{Inverse freeform imaging design}\label{sec: inverse imaging}
In Sec.~\ref{sec: model}, we presented the mathematical model for the design of a freeform imaging system based on inverse methods. The design method depends on the optical map $\boldsymbol{m }:\boldsymbol{\zeta}\mapsto\boldsymbol{y}$ given by Eq.~(\ref{eq: final 3D polar optical map}), which in turn depends on the value of the constant $k.$ We saw that $|k|=\sqrt{k^2}$, where $k^2$ is the ratio of the energy distributions at the source and target of the optical system. Therefore, an imaging system can be computed if we determine the value of the constant $k^2$.

The value of the constant $k^2$ can be calculated using the energy balance given by Eq.~(\ref{eq: 3D energy cons phase space 2}), given the source and target domains, $Q_{\rm{s}}$  and $P_{\rm{t}}$, respectively. A specific choice for these domains, together with the optical map for imaging, enables us to design inverse freeform imaging reflectors using inverse methods from nonimaging optics.

The choice of domains means that all energy from a given source is concentrated on a specified target. This specifies the spread of the reflected light rays as they reach the target plane. For imaging system design, the goal is to minimize aberrations. Thus, choosing a target domain that produces minimal aberrations for a given source domain is essential. 

The Schwarzschild telescope is a double-reflector parallel-to-point imaging system, consisting of reflectors defined as conic sections. It is traditionally known to minimize third-order aberrations \cite{korsch}. The source and target domains corresponding to this classical design are optimal choices for $Q_{\rm{s}}$ and $P_{\rm{t}}$, respectively, for our freeform design. In conventional imaging design methods, classical designs are used as starting points for optimization. By choosing input parameters based on the classical design, we adopt a similar approach.  In Sec.~\ref{sec: results}, we will show that the Schwarzschild telescope does not produce third-order aberrations.

Next, we transform the source $\boldsymbol{q}_{\rm{s}}=\boldsymbol{x}$ to polar coordinates and target momentum $\boldsymbol{p}_{\rm{t}}$ to the stereographic projection $\boldsymbol{y}$ as given by Eq.~(\ref{eq: st proj 3D}). We determine the Jacobian 
\begin{equation}
\det\left(\frac{\partial \boldsymbol{p}_{\rm{t}}}{\partial \boldsymbol{y}}\right)= \frac{4(1-\left|\boldsymbol{y}\right|^2)}{(1+\left|\boldsymbol{y}\right|^2)^3}.
\end{equation}
We transform the stereographic projection $\boldsymbol{y}$ to polar coordinates $(R,\Theta)$. Then, the conservation of energy in Eq.~(\ref{eq: 3D energy cons phase space 2}) results in 
 \begin{equation}\label{eq: 3D energy cons polar}
 	\int_{{\mathcal{A}}} f(r,\vartheta)\,r\,\mathrm{d}r\,\mathrm{d}\vartheta=	\int_{\boldsymbol{{m}}(\mathcal{A})}  g(R,\Theta)\frac{4R(1-R^2)}{(1+R^2)^3}\,\mathrm{d}R\,\mathrm{d}\Theta.
 \end{equation}
where $\mathcal{A}\subset\omega$ and $\boldsymbol{m}(\mathcal{A})\subset\rho$. 
Using Eq.~(\ref{eq: 3D energy cons polar}), we can determine the value of the $f/g=k^2$, for uniform source and target distributions. 

We consider a circular source domain mapped to a circular stereographic target domain, which means that the angles $\vartheta,\Theta\in[0,2\pi]$. For a given source radius $r$, the target radius $R$ can be determined by ray tracing the Schwarzschild telescope. Consider that the total energy at the source is equal to 1. Suppose the source domain has a radius $r$. Then, the source energy distribution $f=1/(\pi r^2)$. The energy balance in Eq.~(\ref{eq: 3D energy cons polar}) is used to determine the uniform target distribution $g$
\begin{equation}
g=\left(2\pi\int_{0}^{R}\frac{4R(1-R^2)}{(1+R^2)^3} \mathrm{d}R\right)^{-1}\\
\end{equation}
Consequently the constant $k^2$ is given by
\begin{equation}
k^2=\frac{f}{g}=\frac{4}{ r^2}\frac{R^2}{(1+R^2)^2}.
\end{equation}

We discuss the choice of sign for $k$ in the phase-space optical map $\boldsymbol{\tilde{m}}$ given by Eq.~(\ref{eq: final 3D linear phase space map}). We ray trace the Schwarzschild telescope and focus on a set of rays generated from a source domain with radius in the interval $r\in(0,a)$, for $0<a<r_{\max}$, and angular coordinate $\vartheta=0^\circ$. These rays have the source coordinates $\boldsymbol{q}_{\rm{s}}=(q_{1\rm{s}},q_{2\rm{s}})^{\rm{T}}$, where $q_{1\rm{s}}=a>0$ and $q_{2\rm{s}}=0$. We observe that the target momentum coordinates of these rays are $\boldsymbol{p}_{\rm{t}}=(p_{1\rm{t}},p_{2\rm{t}})^{\rm{T}}$, with $p_{1\rm{t}}<0$ and $p_{2\rm{t}}=0$. For $q_{1\rm{s}}=a<0$ and $q_{2\rm{s}}=0$, we obtain $p_{1\rm{t}}>0$ and $p_{2\rm{t}}=0$. This implies that  Eq.~(\ref{eq: final 3D linear phase space map}) holds for $k<0$.

In Sec.~{\ref{sec: optical map for imaging}}-{\ref{sec: freeform reflectors}}, we saw that the freeform imaging reflectors are computed by substituting the polar optical map $\boldsymbol{y}=\boldsymbol{m}(\boldsymbol{\zeta})$ from Eq.~(\ref{eq: final 3D polar optical map}) in the cost function formulation given by Eq.~(\ref{eq: 3D cost fxn polar}). The polar optical map $\boldsymbol{m}$ is a transformation of the phase-space optical map $\boldsymbol{\tilde{m}},$ and depends on the same constant $k$ that determines $\boldsymbol{\tilde{m}}$. Thus, $k<0$ also holds for the polar optical map $\boldsymbol{m}$.

\section{Numerical results}\label{sec: results}

We compute the inverse freeform reflectors using the design method presented in Sec.~\ref{sec: model}-\ref{sec: inverse imaging}. We evaluate the performance of the classical design, i.e., Schwarzschild telescope, and the inverse freeform design by comparing the root-mean-square (RMS) spot sizes of the images produced by on-axis and off-axis parallel beams generated from a circular source of a specified radius. In this section, we first discuss ray tracing and then present an example of an imaging system computed using our method.

The RMS spot size $\mathcal{W}$ \cite{RMSspotRadius} is defined as
\begin{equation}\label{eq: 3D rms spot size defn}
\mathcal{W}=\sqrt{\frac{1}{n}\sum_{i=1}^{n}\left|(\boldsymbol{Y}_{\rm{t}})_i-\boldsymbol{\overline{Y}}_{\rm{t}}\right|^2},
\end{equation}
where $\boldsymbol{Y}_{\rm{t}}$ is the point where a ray hits the target plane $z=0$, and $\boldsymbol{\overline{Y}}_{\rm{t}}$ denotes the mean of the position coordinates.
To compute an accurate estimation of the RMS spot size $\mathcal{W}$ corresponding to a parallel beam, we need to precisely calculate the landing position $\boldsymbol{Y}_{\rm{t}}$ of each light ray. The path of a ray is traced by computing the point of intersection on each reflector and subsequently finding the reflected direction using the law of reflection. The vectorial law of reflection \cite[p.~22-24]{lotte} is given by $\boldsymbol{\hat{t}}=\boldsymbol{\hat{s}}-2(\boldsymbol{\hat{s}}\boldsymbol{\cdot}\boldsymbol{\hat{n}})\boldsymbol{\hat{n}}$, where $\boldsymbol{\hat{s}}$ is the direction of incidence, $\boldsymbol{\hat{t}}$ is the reflected direction, and $\boldsymbol{\hat{n}}$ denotes the normal at the point where a ray hits the reflector. This implies that the accuracy of computing $\boldsymbol{Y}_{t}$ depends on the accuracy of the normals, which means that a precise calculation of the first derivatives at the points where rays hit the reflectors is essential.

In our proposed method, reflectors are not given by analytic functions but are computed as discrete point clouds. Thus, our goal is to accurately determine the normals at the reflector surfaces from the data points. Interpolation methods require a large number of points to precisely calculate the normals. However, this would make ray tracing computationally expensive. Hence, we adopt a local approximation method called B-spline quasi-interpolation \cite[Ch.~4]{QI}, that smoothly calculates the derivatives using the neighboring data points only. The detailed construction of this approximation scheme for ray tracing of 2D optical systems was presented in \cite{sanjana2D}. We discuss the generalization to 3D ray tracing in \ref{sec: raytracing}. This scheme is used throughout our ray tracer to determine the surface normals. All computations are done using MATLAB. The intersection points on the reflectors are calculated using the bisection method \cite[p.~146]{bisection}. The stopping criteria were defined by a tolerance of $10^{-10}$, meaning the algorithm terminated when the absolute difference between successive midpoints was less than this value. Additionally, a maximum of $800$ iterations was set to prevent infinite loops in cases of non-convergence.

We present a numerical example to compute the freeform reflectors in a parallel-to-point imaging system, see Fig.~\ref{fig: 3D layout}. We  proceed as follows:
\begin{figure}[h!]
		\centering
				\includegraphics[width=0.5\textwidth]{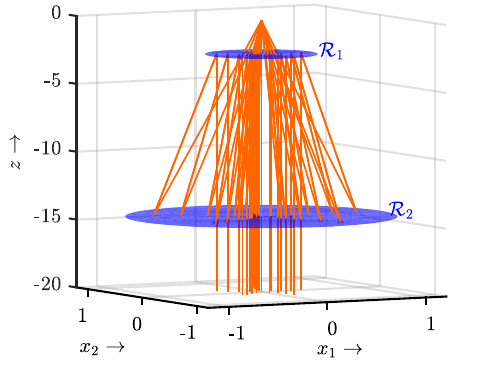}
    \caption{Layout of the optical system.}
        \label{fig: 3D layout}
    \end{figure}
\begin{enumerate}
\item \textit{System layout for inverse design}: We fix the distance $D_{\rm{v}}=12$ between the vertices of both reflectors and the location of the source at $z=-20$. We will choose other parameters later.

\item \textit{System layout for classical design}:  For the Schwarzschild design, we choose the design parameters that are known for the maximum correction of third-order aberrations \cite{korsch}. The shape of each reflector defined on a polar grid is given by
    \begin{equation}\label{eq: 3D schwarz}
    z=\mathcal{R}_{i}(\sigma,\varsigma)=\frac{\sigma^2}{r_{i}+\sqrt{r_{i}-(1+C_{i})\sigma^2}}, \quad i=1,2,
    \end{equation}
    where $\sigma\in\mathbb{R}^{+}$ is the radial distance from origin, $\varsigma\in[0,2\pi]$ is the angular coordinate, $C_i$ are the deformation constants, $r_i$ are the radii of curvature, and subscript $i$ corresponds to the first and second reflectors. Let $f_{\rm{s}}$ denote the the focal length of the system. We choose $D_{\rm{v}}=12,$ and all other parameters are given in Table.~\ref{table: telescope parameters}.

    \begin{table}[h!]
	\renewcommand{\arraystretch}{1.3}
	\setlength{\tabcolsep}{1cm}
	\caption{Parameters for a Schwarzschild telescope.}
	\normalsize
	\label{table: telescope parameters}
 \centering
	\begin{tabular}{|c|c|}
		\hline
        \textbf{Parameter}& \textbf{Value}\\
        \hline
        $r_1$&$-2f_{\rm{s}}\sqrt{2}$ \\
        $r_2$&$-2f_{\rm{s}}\sqrt{2}$\\
      $  D_{\rm{v}}$&$-2f_{\rm{s}}$\\
      		$C_1$&$(1+\sqrt{2})^2$\\
       
        		$C_2$&$(1+\sqrt{2})^{-2}$\\

		\hline
	\end{tabular}
\end{table}
 \begin{figure}[hb]
		\centering
				\includegraphics[width=0.7\textwidth]{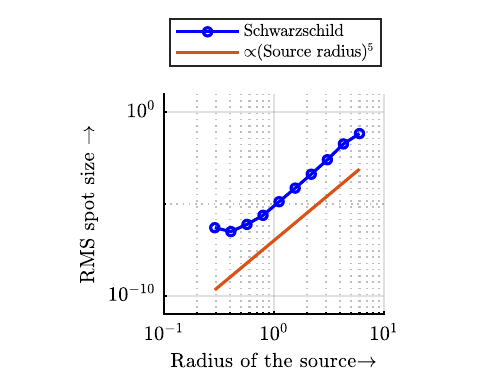}
    \caption{The Schwarzschild telescope has fifth-order aberrations.}
        \label{fig: fifth order}
    \end{figure}

\item \textit{Ray trace classical design}: Using Eq.~(\ref{eq: 3D schwarz}), we calculate both reflectors for the classical design. The first and second reflectors are defined on circular grids with radii equal to $0.5$
 and $1.5$, respectively. A $151\times151$ polar grid is used to discretize the domain. We consider a circular source domain $\omega$ with radius $r_{\max}=0.5$. We ray trace $100$ uniformly generated on-axis rays passing through the classical design and observe that the corresponding circular stereographic target domain $\rho$ has radius $R=0.0417$.
 Furthermore, we verify that this design is limited by fifth-order aberrations, as shown by Fig.~\ref{fig: fifth order}. This justifies that $R$ is an ideal input parameter for inverse freeform design. In Fig.~\ref{fig: fifth order}, the numerical errors dominate for small source radii. 
 
     \begin{figure}[t]
		\centering
   
				\includegraphics[width=0.7\textwidth]{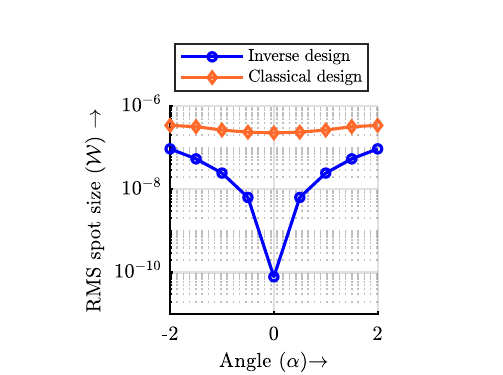}
    
    \caption{Comparison of RMS spot sizes $\mathcal{W}$ for various angles $\alpha$.}
    \label{fig: 3D compare spot size}
    \end{figure}
      \begin{table}[t]
	\renewcommand{\arraystretch}{1.1}
	\caption{RMS spot sizes for different angles.}
	\normalsize
	\label{table: 3D spot sizes}
    \centering
	\begin{tabular}{|c|c|c|}
			\hline
				\textbf{Angle}& $0^{\circ}$& $\pm2^\circ$\\
				\hline
				\textbf{Inverse Design} & $7.417e-11$ & $9.075e-08$ \\
                \textbf{Classical Design} & $1.386e-07$ & $3.461e-07$ \\
				\hline
	\end{tabular}
\end{table}
\item \textit{Compute inverse design}:  From the above source and target domains and for uniform energy distributions $f$ and $g$, the energy balance given in Eq.~(\ref{eq: 3D energy cons polar}) leads to $k=-0.1666$. Next, Eq.~(\ref{eq: 3D optical map polar coordinates}) is used to compute the polar optical map $\boldsymbol{y}=\boldsymbol{m}(\boldsymbol{\zeta})$, where $\boldsymbol{\zeta}$ is discretized with $151\times151$ polar grid points and $r\in[0,0.5]$ and $\vartheta\in[0,2\pi]$. As elaborated in Sec.~\ref{sec: freeform reflectors}, $\tilde{u}_1$ is computed using Eq.~(\ref{eq: 3D u1 ODE}) by specifying the distance between the vertex of the first reflector and the source plane, i.e., $u_0=17.522$. The values for $\tilde{u}_1$ are computed with MATLAB's ODE solver \textit{ode113} with \textit{RelTol}$=10^{-12}$ and \textit{AbsTol}$=10^{-16}$.The shape of the first reflector $\mathcal{R}_1$ is given by Eq.~(\ref{eq: 3D first reflector}). The OPL is chosen equal to that in the classical design, i.e., $V=44$. Eqs.~(\ref{eq: 3D cost fxn polar}) and (\ref{eq: 3D second reflector}) are then used to compute the second reflector $\mathcal{R}_2$.
 \end{enumerate}

The direction of an incident beam can be fully described by two angles in spherical coordinates: a polar angle $\alpha$, which defines the tilt relative to the $z$-axis, and an azimuthal angle, which defines rotation around the $z$-axis. Since we consider a rotationally symmetric system, the behavior of the system is invariant with respect to the azimuthal angle, and depends only on the polar angle $\alpha$. 
On-axis rays correspond to $\alpha=0^{\circ}$. All other rays are off-axis.

 We consider a circular source domain with radius equal to $0.5$. For the classical and inverse designs, we ray trace $100$ rays, with $\alpha \in \{-2^{\circ},-1.5^{\circ},\ldots,2^{\circ}\}$. In Table~\ref{table: 3D spot sizes}, we give the RMS spot sizes corresponding to different $\alpha.$ Fig.~\ref{fig: 3D compare spot size} demonstrates that the inverse freeform design clearly outperforms the classical design for both on-axis and off-axis parallel beams.

   \section{Conclusions and future work}\label{sec: conclusion}
We presented an inverse method for the design of a 3D double-reflector imaging system with a parallel source and a point target. We illustrated that the phase-space optical map for ideal imaging is a scalar multiple of a $2\times2$ identity matrix. We imposed this optical map and derived that for imaging design based on inverse methods, the ratio of energy distributions at the source and target is a constant. Furthermore, the requirement of a classical design that minimizes aberrations is discussed. We demonstrated that inverse methods from nonimaging optics can be extended to design imaging systems.

We tested the performance of our inverse design by comparing the RMS spot sizes of
on-axis and off-axis rays with those produced by the classical design. The inverse design is tailored to produce an aberration-free image for on-axis rays only. The classical design minimizes aberrations for off-axis rays. The inverse design yields reduced RMS spot sizes not only for on-axis rays but also for off-axis rays and outperforms the classical design in both cases. This means that the inverse design is superior to the classical design.

Some possible extensions for future work are to determine and analyze the optical aberrations produced by the inverse design and then minimize them for improved performance. An extensive design and manufacturing tolerance analysis may be required to utilize the proposed freeform design in various applications. The proposed method can be adapted to design folded imaging systems, which can be used in telescopic applications. It might be interesting to see the effect of adding more optical surfaces to the system, as it leads to more degrees of freedom for design.

\appendix
\section{Quasi-interpolation for ray tracing}\label{sec: raytracing}

For ray tracing our imaging system, we aim to precisely approximate both reflector surfaces and the corresponding surface normals from a collection of 3D data points. We construct a 2D quasi-interpolation (QI) scheme for a real-valued function, based on tensor-product spline spaces. Furthermore, we discuss the method to calculate first-order partial derivatives, which are required to determine surface normals. 

Consider a real-valued function $h(x,y)\in \mathcal{B}_d(x,y)$, where $\mathcal{B}_d(x,y)$ denotes a bivariate polynomial space
\begin{equation}
 \hspace{-1cm}   \mathcal{B}_d(x,y)=\{ h: \mathbb{R}^2\to \mathbb{R} \mid h(x,y)=\sum_{m,n}a_{mn}x^my^n,\quad 0\leq m,n\leq d\}.
\end{equation}
Let $B_{r,d,\boldsymbol{\xi}}(x)$ denote the $r^{th}$ B-spline of degree $d$ defined on an open knot sequence $\boldsymbol{\xi}$ in the $x$-direction, where \cite{QIlychee}
\begin{equation}
\hspace{-2cm}\boldsymbol{\xi}:=\{\xi_i\}_{i=1}^{n+d+1}=\{\xi_{1}=\cdots=\xi_{d+1}<\xi_{d+2}\leq\cdots\leq\xi_{n}<\xi_{n+1}=\cdots=\xi_{n+d+1}\}.
\end{equation}
Similarly, we define B-splines $B_{s,d,\boldsymbol{\mu}}(y)$ in the $y$-direction, where $\boldsymbol{\mu}:=\{\mu_j\}_{j=1}^{n+d+1}$ is an open knot vector with the same structure as $\boldsymbol{\xi}$.
Consider that function values $h(\bar{x},\bar{y})$ are known at grid points $(\bar{x},\bar{y})$, where $\bar{x}=\{\bar{x}_i\}$, $\bar{y}=\{\bar{y}_j\}$ and $(\bar{x}_i,\bar{y}_j)$ are discrete grid points. Hereafter, the bar symbol $(\,\bar{}\,)$ will be used to denote the points where function values are known.

We discuss the implementation of the B-spline QI method in 2D using the tensor-product splines. The tensor-product spline approximation \cite[Ch.~7]{QIlecturenotes} is given by
\begin{equation}\label{eq: 2D QI}
h(x,y)\approx\sum_{s=1}^{n}\sum_{r=1}^{n} \lambda_{rs}B_{r,d,\boldsymbol{\xi}}(x)B_{s,d,\boldsymbol{\mu}}(y).
\end{equation}
We construct the weights $\lambda_{rs}$ using the QI method, a local approximation scheme in which the weights depend on the values of the function $h$ at grid points in the neighborhood of the point $(x,y)$. We will discuss details in the later parts.

The sets $\{B_{r,d,\boldsymbol{\xi}}(x)\}_{r=1}^{n}$ and $\{B_{s,d,\boldsymbol{\mu}}(y)\}_{s=1}^{n}$ form the basis that spans the univariate spline spaces, $S_1$ in the $x$-direction, and $S_2$ in the $y$-direction, respectively. Thus, the functions 
$\{B_{r,d,\boldsymbol{\xi}}(x)B_{s,d,\boldsymbol{\mu}}(y)\}_{r,s}$ form a basis for the tensor-product space $S_1\otimes S_2$. Eq.~(\ref{eq: 2D QI}) means that the approximation $h$ lies in the bivariate spline space $S_1\otimes S_2$, which consists of independent spline spaces in each direction. This implies that the $x$- and $y$-directions can be treated independently during the approximation of the function $h$, and its first-order partial derivatives, $\mathrm{D}_{x}h$ and $\mathrm{D}_{y}h$. 

\textbf{Approximation of the function $h$:} For any real-valued 1D function $h^*\in\mathcal{P}_d$, where $\mathcal{P}_d$ denotes the space of polynomials of degree at most $d$, a B-spline quasi-interpolant \cite[Chap.~2]{QI} is defined as an operator of the form
\begin{equation}\label{eq: QI defn}
    	Q_d\,[h^*](x)=\sum_{p=1}^{n}\lambda_{p}(h^*)B_{p,d,\boldsymbol{\xi}}(x),
\end{equation}
where $\lambda_{p}(h^*)$ are coefficients defined as a linear combination of some known values of $h^*$ at some points in the neighborhood of the support of $B_{p,d,\boldsymbol{\xi}}(x)$. A QI scheme is exact on the space of polynomials of degree at most $d$, implying that using Eq.~(\ref{eq: QI defn}), polynomials can be locally approximated as a linear combination of B-spline basis functions. The following QI approximation holds in 1D
\begin{equation}\label{eq: 1D QI}
    h^*(x)\approx\sum_{p=1}^{n}\lambda_{p}[h^*](\bar{x})B_{p,d,\boldsymbol{\xi}}(x),
\end{equation}
where the weights $\lambda_{p}[h^*](\bar{x}):=\lambda_{p}(h^*(\bar{x}))$ represent functionals that depend on the known values of $h^*$ at grid points $\bar{x}$. We will consistently use this notation throughout to indicate functional dependence, with the argument specifying the relevant grid point.

The separable structure in Eq.~(\ref{eq: 2D QI}) allows a sequential construction of the 2D approximation scheme based on the above 1D QI scheme. First, QI is applied in the $x$-direction by fixing $\bar{y}_j$ and computing a univariate intermediate approximation $\tilde{h}$. Second, the intermediate representation $\tilde{h}$ is utilized for approximating the $y$-direction using the basis functions $\{B_{s,d,\boldsymbol{\mu}}(y)\}_{s=1}^{n}$. The procedure is as follows.
\begin{enumerate}
\item For each fixed $\bar{y}_j$, Eq.~(\ref{eq: 1D QI}) is used to determine the 1D QI approximation (in the $x$-direction) 
\begin{equation}\label{eq: 2D QI tilde{h}}
    \tilde{h}_j(x,\bar{y}_j)\approx\sum_{r=1}^{n} a_r[h](\bar{x},\bar{y}_j)B_{r,d,\boldsymbol{\xi}}(x),
\end{equation}
where $a_r$ are the weights depending on the known function values $h(\bar{x},\bar{y}_j)$. The method to calculate these weights for a symmetric 1D QI scheme employing cubic B-splines is presented in \cite{sanjana2D}. All weights associated with QI in 1D are determined using the same method in the sequel.
\item Let $\tilde{h}(x,\bar{y})$ denote the approximation of the function $h$ in the $x$-direction, evaluated for $\bar{y}$. We use the function values $\tilde{h}(x,\bar{y})$ and use QI in the $y$-direction to obtain the 2D approximation 
\begin{equation}
h(x,y)\approx\sum_{s=1}^{n}b_s[\tilde{h}](x,\bar{y})B_{s,d,\boldsymbol{\mu}}(y),
\end{equation}
\end{enumerate}

\textbf{Approximation of the first-order partial derivatives}: In 1D, the derivative $\mathrm{D}h^{*}$ of a function $h^*\in\mathcal{P}_d$ is given by
\begin{equation}\label{eq: 1D QI derivative}
    \mathrm{D}h^{*}(x)\approx\sum_{p=2}^{n} c_p[h]^*(\bar{x})B_{p,d-1,\boldsymbol{\xi}}(x), \quad c_p=d\frac{\lambda_p[h^*](\bar{x})-\lambda_{p-1}[h^*](\bar{x})}{\xi_{p+d}-\xi_d}.
\end{equation}
We use a similar sequential procedure for approximating the first-order partial derivatives as used for approximating the function $h$.

\begin{itemize}
\item \textbf{Approximation of} $\mathrm{D}_{x}\,h$: For each fixed $\bar{y}_j$, we use Eq.~(\ref{eq: 1D QI derivative}) to obtain the partial derivative with respect to $x$
\begin{equation}
  \hspace{-2.2cm}  \mathrm{D}_{x}\tilde{h}_j(x,\bar{y}_j)\approx\sum_{r=2}^{n} a'_r[h](\bar{x},\bar{y}_j)B_{r,d-1,\boldsymbol{\xi}}(x), \quad a'_r=d\frac{a_r[h](\bar{x},\bar{y}_j)-a_{r-1}[h](\bar{x},\bar{y}_j)}{\xi_{r+d}-\xi_d}.
\end{equation}
Next, we use $\mathrm{D}_{x}\tilde{h}(x,\bar{y})$ to obtain
\begin{equation}
 \mathrm{D}_{x}h(x,y)\approx\sum_{s=1}^{n} b_s[\mathrm{D}_{x}\tilde{h}](x,\bar{y})B_{s,d-1,\boldsymbol{\mu}}(y).
\end{equation}
\item \textbf{Approximation of} $\mathrm{D}_{y}\,h$: We use Eq.~(\ref{eq: 2D QI tilde{h}}) to determine the intermediate approximation $\tilde{h}(x,\bar{y}).$  Then, using Eq.~(\ref{eq: 1D QI derivative}), the partial derivative with respect to $y$ reads as
\begin{equation}
\hspace{-2cm}   \mathrm{D}_{y}h(x,y)\approx\sum_{s=2}^{n} b'_r[\tilde{h}](x,\bar{y})B_{s,d-1,\mathrm{\mu}}(y), \quad b'_r=d\frac{b_r[\tilde{h}](x,\bar{y})-b_{r-1}[\tilde{h}](x,\bar{y})}{\mu_{s+d}-\mu_d}.
\end{equation}

\end{itemize}

\section*{Funding} This research is supported by Topconsortium voor Kennis en Innovatie (TKI program ``Photolitho MCS" (TKI-HTSM 19.0162)).

\section*{References}
\bibliographystyle{iopart-num}  
\bibliography{sample}

\end{document}